# *Nanofilters for Optical Nanocircuits*

*Andrea Alù, Michael Young, and Nader Engheta[*]*

*University of Pennsylvania*

*Department of Electrical and Systems Engineering*

*Philadelphia, Pennsylvania 19104, U.S.A.*

**Abstract**

We theoretically and numerically study the design of optical 'lumped' nanofiltering devices in the framework of our recently proposed paradigm for optical nanocircuits. In particular, we present the design of basic filtering elements, such as low-pass, band-pass, stop-band and high-pass 'lumped' nanofilters, for use in optical nanocircuits together with more complex designs, such as multi-zero or multi-pole nanofilters, to work at THz, infrared and optical frequencies. Following the nanocircuit theory, we show how it is possible to design such complex frequency responses by simple rules, similar to RF circuit design, and we compare the frequency response of these optical nanofilters with classic filters in RF circuits. These results may provide a theoretical foundation for fabricating nanofilters in optical lumped nanocircuit devices.

*1.   Introduction*

The importance of transplanting the classical circuit concepts into optical frequencies is based on the possibility of squeezing circuit functionalities (e.g., filtering, waveguiding, multiplexing…) in subwavelength regions of space, and on correspondingly increasing the operating frequency with several orders of magnitude. Moreover, nowadays the interest in combining optical guiding devices, as optical interconnects, with micro- and nanoelectronic circuits is high (see, e.g., [1]), since it is not still possible to perform all the classic circuit operations in the optical domain. Introducing new paradigms and feasible methods to bring more circuit functionalities into the optical domain would represent an important advance in nanoelectronics technology.

In a recent letter [2], we have introduced and discussed the fundamental concepts for developing a novel paradigm for optical nanocircuits, with the aim to extend classic circuit concepts, commonly available at RF and lower frequencies, to higher frequencies and in particular to the optical domain. Specifically, we have discussed in [2] and in following papers [3]-[9] how a proper combination of plasmonic and non-plasmonic nanoparticles may constitute a complex nanocircuit at infrared and optical frequencies, for which the conventional lumped circuit elements are not available in a conventional way. After introducing the nanocircuit concepts for isolated nanocircuit elements [2], and after having applied them to model infinite stacks of nanoelements to design nanotransmission lines and nanomaterials [3]-[5], we have been interested in analyzing in details how the connections and interactions among the individual nanoelements may be modeled and designed in a complex optical *nanocircuit board* with functionalities corresponding to those of a classic microwave circuit. In [6]-[8] we have indeed analyzed basic coupling issues among individual nanocircuit elements and we have proposed the use of ε-near-zero (ENZ) and ε-very large (EVL) materials, with ε being the local material permittivity, in order to operate as "insulators" and "connectors", respectively, for proper interconnections among the different elements of a nanocircuit systems. In [9] we have also proposed a specific geometry to model a "shorting wire" at optical frequencies, in order to connect relatively "distant" sections of a nanocircuit board. The interest in interpreting the light interaction with plasmonic materials in terms of circuit models has been reported in the literature earlier (see, e.g., [10]-[13]). However, the novelty of our approach [2] consists in the ability to tailor and design nanoparticles to act as "lumped" circuit elements (e.g., nanoinductors, nanocapacitors, and nanoresistors) with given lumped impedances at optical frequencies. From this concept, one will then be able to *quantitatively* design and synthesize more complex nanocircuits and nanosystems at optical frequencies, using collections of nanoparticles acting as "lumped" nanocircuit elements. As we discussed in [2], in fact, a complete equivalence between RF circuits and optical nanocircuits relies on the analogy between the local conduction current density ($\mathbf{J}_c = \sigma \mathbf{E}$) and the local displacement current density ($\mathbf{J}_d = -i\omega\varepsilon\mathbf{E}$), with $\sigma$ and $\varepsilon$ being the local material conductivity and permittivity, respectively, and $\mathbf{E}$ the local electric field. The proportionality of $\mathbf{J}_d$ with frequency, added to the usual frequency dispersion of the material permittivity $\varepsilon$, which generally follows the Drude-Lorenz dispersion models [14]-[15], implies that at higher frequencies the role of conduction current in optical nanocircuits is overtaken by the role of $\mathbf{J}_d$ [2]. This explains why the role of conductivity $\sigma$ in low-frequency circuits is "replaced" by the local permittivity $\varepsilon$ in optical nanocircuits. As we mentioned in [2], it follows that plasmonic (negative ε) and dielectric (positive ε) nanoparticles may constitute, respectively, lumped nanoinductors and nanocapacitors, whereas ENZ or EVL materials in this paradigm correspond, respectively, to "poor conductors", or nanoinsulators, and "good conductors", or nanoconnectors [6]. With these tools, it is possible to envision more complex interconnections among the different lumped nanocircuit elements, or even nanomodules, of a nanocircuit board, in order to realize complex frequency responses at infrared or optical frequencies.

Following this same analogy between classic circuit theory and these optical nanocircuits, we go one step further to present a complete analysis and some designs for nanoscale filtering modules at optical frequencies. In this scenario, we apply the familiar concepts of circuit and filter theories, in order to design first and second-order nanofilters (in Section 2) and higher-order nanofilters (in Section 3) with complex frequency responses by combining and juxtaposing up to four reactive nanoelements. Our results seek to pave the way for the future design of functional lumped optical nanocircuits at frequencies where the classic RF lumped circuit elements are not available.

## 2. *Basic filtering elements: first- and second-order optical nanofilter design*

The fundamentals of filter design in low-frequency circuits are based on the frequency responses of lumped capacitors and inductors [16] and their combinations. For instance, a standard low-pass or high-pass filter is based on the voltage extracted from a resistor-capacitor (RC) pair or a resistor-inductor (RL) pair, respectively. An LC pair, due to its inherent resonant properties, constitutes the fundamental pass-band or stop-band filter. In this section we consider these basic elements as the starting point for our lumped optical nanofilter design. In particular, in the following we will consider nanocircuits embedded in a parallel-plate waveguide, which guides the input optical excitation. Following the theoretical results reported in [2], all the nanocircuit elements will be subwavelength in size at the frequency of interest, in order to allow them to function as lumped elements. For the sake of simplicity, the structures analyzed here are two-dimensional (2D), and the cross sections of all of our nanostructures (i.e., nanorods) studied here are small compared with the operating wavelengths. Extension to a fully three-dimensional (3D) geometry would not change the main concepts presented here. The numerical results are fully dynamic, extracted from full-wave simulations by commercially available finite-integration time-domain software [17], in some examples also compared to finite-element software [18], and consider realistic dispersion and reasonable losses for the involved materials. In all of the examples, the results are compared with what is expected from an analogous design in a low-frequency RF circuit, considering the quantitative values of the involved nanocircuit elements as predicted by nanocircuit theory [2], and in each case a sketch of the related circuit model is shown. As already mentioned, the functionalities of the lumped nanofilters considered here have been evaluated inside a parallel-plate waveguide, which may correspond at optical frequencies to a metal-insulator-metal waveguide (see, e.g., [3], [19]). Analogous responses are expected when the same nanocircuit elements are employed in different setups, e.g., when printed on a substrate or when utilized as "loads" in optical nanoantennas. Such geometries, which may be closer to a practical realization at optical frequencies, are currently under investigation by our group, but they go beyond the interest of the present work, in which we aim to fully characterize the nanofiltering response of proper combinations of nanocircuit elements.

Since the goal of the present work is to determine the filtering transfer functions of collections of nanoelements, here we assume that the host waveguide is made of perfectly conducting walls and thus a transverse electromagnetic (TEM) mode is launched at the input of this waveguide. Therefore, the frequency response of the "empty" waveguide is assumed to be uniform and flat. In a realistic scenario, the metallic walls of the waveguide are dispersive and thus the waveguide itself exhibit certain well-known (relatively weak) dispersion, which will 'modulate' the filtering function of the nanofilters. This additional dispersion can of course be taken into account in any realistic design of such nanofilters.

*a) Low-Pass optical nanofilter: parallel RC combination*

As a first example, Fig. 1 reports the design of a low-pass nanofilter. The geometry is sketched in the inset, and is represented inside the cross-section of a parallel-plate waveguide (with the impinging electric field oriented from bottom to top of the figure) in air (permittivity $\varepsilon_0$). The low-pass nanofilter is formed as an "RC parallel combination", as the circuit sketch reports in the inset of the figure, where the parallel resistance is provided by the intrinsic impedance of the waveguide $\eta$. The "nanocapacitor" C is formed by a dielectric nanorod of permittivity $\varepsilon_C = 16\varepsilon_0$ (close to silicon-based materials at infrared-optical frequencies) with base $d = 20\,\text{nm}$ and height $h = 10\,\text{nm}$. The geometrical and electromagnetic parameters have been chosen as possible design values, but the concepts outlined here are in principle scalable to different frequencies or different material parameters.

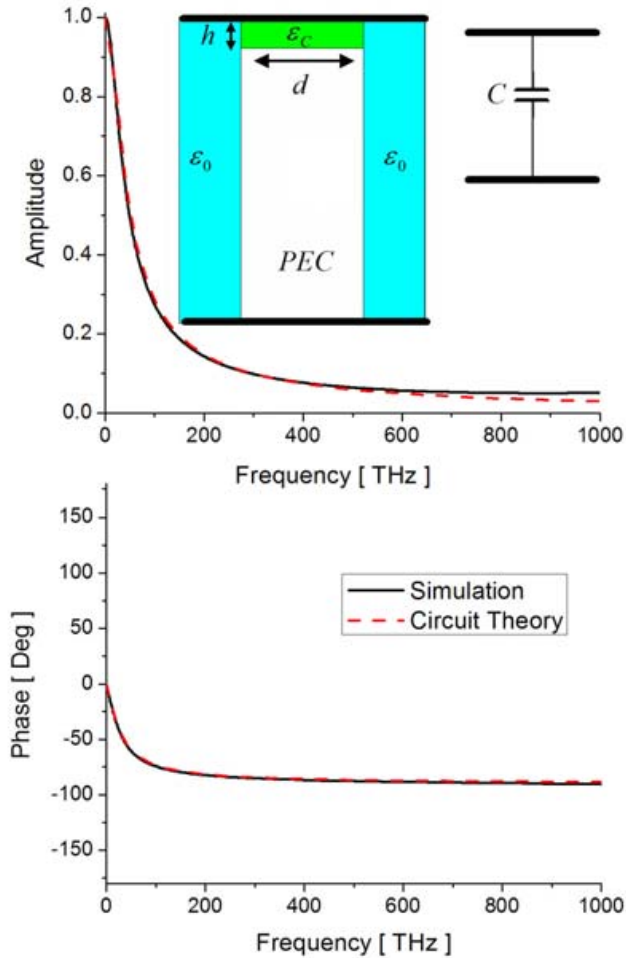

Figure 1 - Transfer function (amplitude and phase) for the basic low-pass optical nanofilter made of a dielectric nanorod in a waveguide depicted in the inset. The black solid line refers to full-wave simulations, whereas the red dashed line corresponds to circuit-theory calculations for the circuit model sketched in the inset.

The parallel-plate waveguide has a total height $h_{wg} = 100$ nm and it is closed on top and bottom by perfect electric conductors, which may well represent some impenetrable metal with sufficiently negative permittivity in the frequency regime of interest. The "nanocapacitor" is connected to the PEC protrusion attached to the bottom of the waveguide. The capacitance per unit length of the nanoparticle may be evaluated following [2], yielding $C = \varepsilon_c d / h = 0.283 \dfrac{nfarad}{m}$. The parallel "resistance" R of the nanofilter, as mentioned above, is provided by the characteristic impedance of the waveguide, which in this geometry is given by: $\eta = \sqrt{\mu_0/\varepsilon_0}\, h_{wg} = 18.85\,\mu\Omega \times m$, with $\mu_0$ being the permeability of free space.

The extracted transfer function for this basic optical nanofilter is shown in Fig. 1, in amplitude and phase (black lines) and they are compared with those obtained through the circuit theory for the same values of R and C. The agreement between the lines is quite good, despite the drastic abruption in the waveguide cross section and the dynamic nature of the problem, which does not consider any "nanoconnector" or "nanoinsulator" around the nanoelement. This is due to the small dimensions of the nanocircuit with respect to the wavelength of operation and to the absence of a resonant response, which ensure the "lumped" behavior of the nanocircuit over the frequency range of interest. A very minor discrepancy in the amplitude and phase of the extracted transfer function is seen at higher frequencies, when the nanocapacitor size becomes closer to the wavelength of operation. The amplitude of the nanofilter rapidly decays in frequency, and the phase passes from 0 to -90 degrees, consistent with the functionality of an RC-parallel low-pass filter. It should be noted that the PEC step in the waveguide was inserted to increase the value of the small capacitance with respect to the characteristic impedance of the waveguide, in order to fine tune the cut-off frequency of the nanofilter at the frequency of interest. This provides an interesting design tool to tailor the frequency response of the filter for the designer's needs.

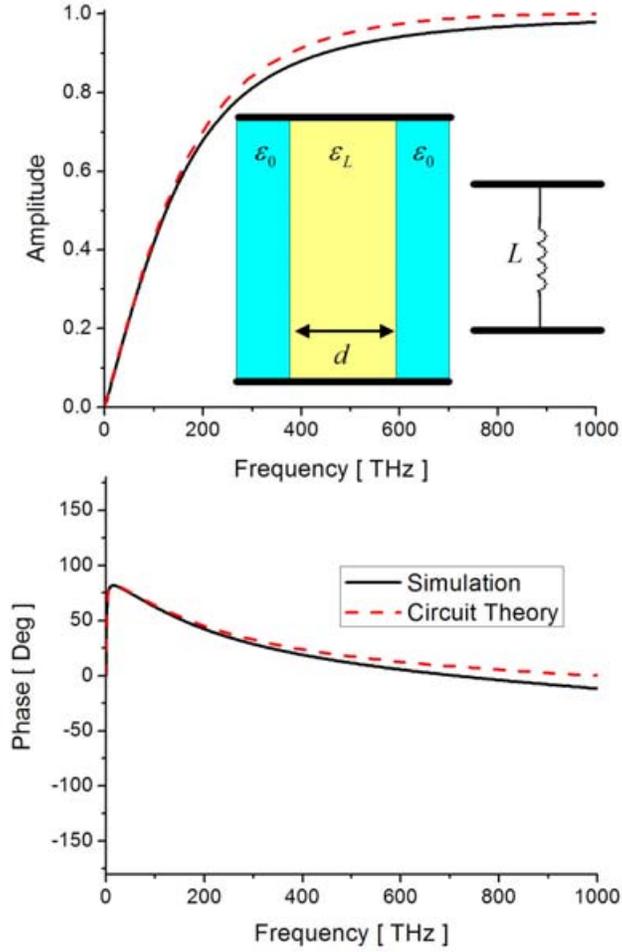

Figure 2 - Transfer function (amplitude and phase) for the basic high-pass optical nanofilter formed by a plasmonic nanorod in a waveguide depicted in the inset. The black solid line refers to full-wave simulations, whereas the red dashed line corresponds to the circuit-theory calculations for the circuit model sketched in the inset.

*b) High-Pass optical nanofilter: parallel RL combination*

A simple high-pass filter is depicted in Fig. 2. It is formed as an "RL parallel combination", where the "nanoinductor" L, as first suggested in [2], is a plasmonic nanorod with cross sectional dimensions $(h_{wg} = 100\,\text{nm}) \times (d = 20\,nm)$. This nanorod has a permittivity following a Drude model, i.e.:

$$\varepsilon_L = \varepsilon_0 \left(1 - \frac{f_p^2}{f(f+i\gamma)}\right), \quad (1)$$

with $f_p = 10^3$ THz and $\gamma = 1$ THz, which models reasonable dispersion and losses for a plasmonic material at infrared/optical frequencies. Also in this case the "resistance" R is provided by the characteristic impedance of the waveguide. Again, full-wave simulations and circuit predictions coincide very well, and the high-pass functionality of this nanofilter is evident from the plots in Fig. 2.

It should be noted that in this example the nanoinductor has some intrinsic small losses included in it, due to the absorption of the plasmonic material. Moreover, the value of the associated equivalent inductance per unit length of this nanorod is inherently dispersive, generally changing with frequency following the formula $L = -\dfrac{h_{wg}}{\omega^2 \varepsilon_L(\omega) d}$ [2]. These aspects are fully taken into account in the nanocircuit model. The presence of losses in the material explains why the phase of the transfer function goes abruptly to zero at very low frequencies. The smooth transition of the phase from +90 deg to zero is typical of a high-pass RL filter. As an aside, we note that for frequencies sufficiently lower than $f_p$, but still greater than $\gamma$, the value of $\varepsilon_L$ becomes approximately $\varepsilon_L \simeq -\dfrac{\omega_p^2}{\omega^2}$, which results in an approximately constant effective inductance per unit length $L = \dfrac{h_{wg}}{\varepsilon_o \omega_p^2 d}$.

*c) Pass-Band optical nanofilter: parallel RLC resonance*

Figure 3 reports the design of a "parallel RLC" optical nanofilter working as a pass-band device. The nanofilter was designed by juxtaposing two nanorods with cross sectional dimensions $(d = 20\,\text{nm}) \times (h = 10\,\text{nm})$. The two nanoelements have, respectively, permittivity $\varepsilon_C = 14.15\varepsilon_0$, which corresponds to the permittivity of silicon at optical frequencies [20] and $\varepsilon_L = \left(5 - \dfrac{f_p^2}{f(f+i\gamma)}\right)\varepsilon_0$, where $f_p = 2175\,\text{THz}$ and $\gamma = 4.35\,\text{THz}$, which models silver at optical frequencies [21]. Once again, the two nanorods have been embedded in the waveguide by positioning them on top of a step made of an impenetrable material (the same that composes the parallel plates on top and bottom of the waveguide). The geometry in the figure shows the distribution of electric field in the waveguide at the resonance frequency, which occurs at $420\,\text{THz}$.

Consistent with our analysis in [7], the two nanocircuit elements are in parallel due to the orientation of the impinging optical electric field (parallel to the common interface) and the optical potential drop from the top to the bottom of the two nanoelements is indeed the same. The circuit equivalent is shown in the inset of Fig. 3, and the corresponding transfer function is compared with the simulated one in the plots. The simulated and theoretical results match reasonably well, both in amplitude and phase, despite a small shift in the resonance frequency, due to the evident mismatch between the waveguide cross section and the nanoparticle cross-section at the resonance. Since the two nanorods have the same cross-sectional dimensions, from the circuit theory the resonance is expected to occur when the two material permittivities have the same magnitudes but opposite signs, i.e., around $500\,\text{THz}$. The actual resonance is shifted down, possibly due to the local fringing fields at the waveguide abruption that slightly increase the value of C. It is also evident how the presence of resonances in the nanocircuit slightly detunes the expected transfer function from the simulated one, when compared with the previous examples. Nevertheless, the agreement between the two lines is reasonably good, and with a proper optimization, or the use of nanoinsulators and nanoconnectors, this may be improved.

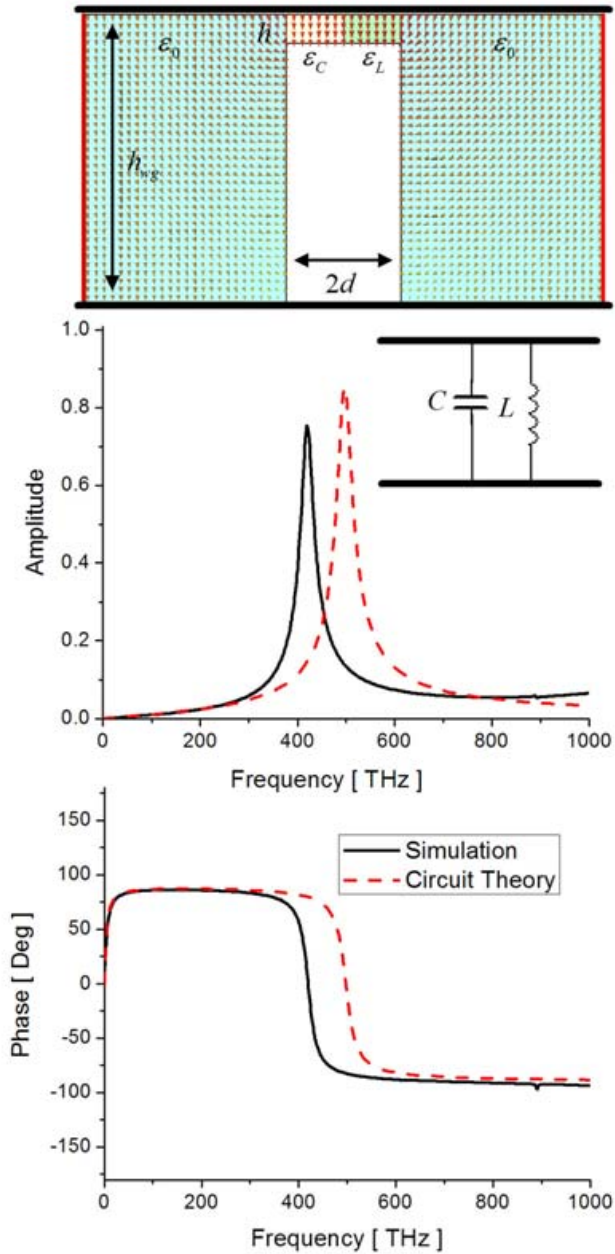

Figure 3 - Transfer function (amplitude and phase) and electric field distribution at the resonance for an optical pass-band nanofilter formed by two nanorods juxtaposed in parallel in a waveguide, one made of silicon and the other made of silver.

As in the low-pass nanofilter case, the choice of reducing the height $h$ of the nanoelements with respect to the height of the waveguide $h_{wg}$ is motivated by the possibility of increasing the value of C and simultaneously reducing the value of L, following the previous formulas. In this case, the Q-factor of the "parallel-RLC resonance", given by the standard expression [16]:

$$Q = R\sqrt{\frac{C}{L}}, \qquad (2)$$

increases, and therefore the pass-band becomes narrow around the frequency of interest.

If we considered $h = h_{wg}$, we would indeed get the advantage of no abruption in the waveguide section, and therefore a better agreement between the circuit theory and full-wave simulations. However, in this case the wider bandwidth of operation of the filter would make this optical nanofilter less useful at optical frequencies.

As suggested in [6], the use of proper ENZ nanoinsulators on the "front" and the "back" of the nanocircuit, in order to confine the undesired displacement current leakage out of the resonant pair in this case for which $h < h_{wg}$, might improve the agreement between circuit theory and simulations around the resonance frequency.

The material permittivities, dimensions of the nanorods' cross sections, and the height of the step represent degrees of freedom in the design to achieve the required frequency response for the pass-band filter, both in terms of central frequency and bandwidth of operation.

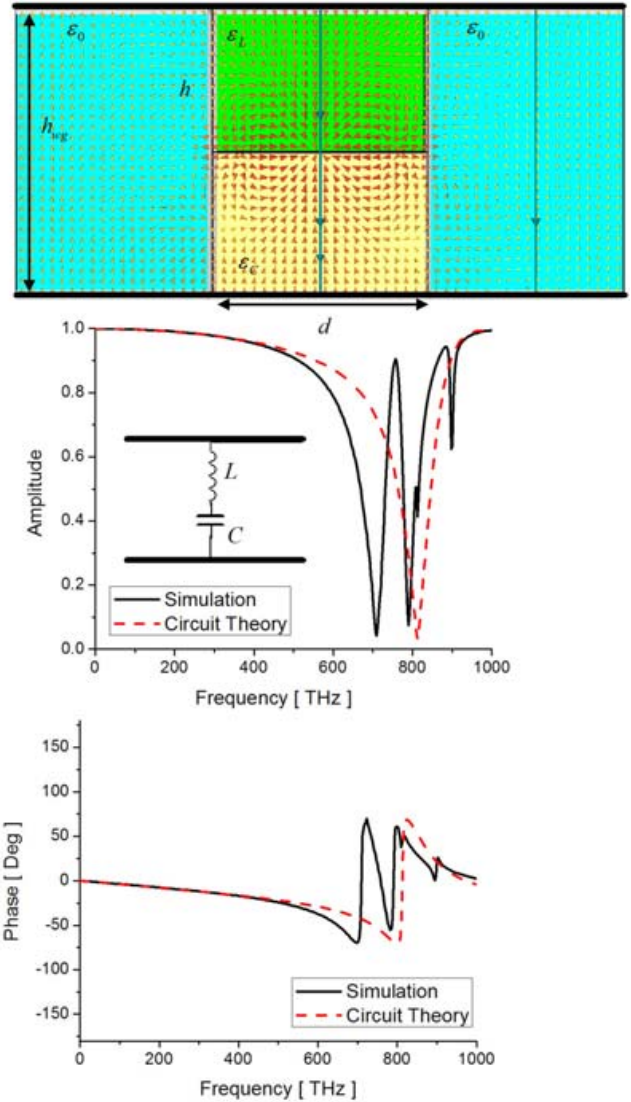

Figure 4 - Electric field distribution at resonance and the transfer function for a stop-band nanofilter formed by two nanorods juxtaposed in series in a waveguide, one made of and the other made of silver.

*d) Stop-Band optical nanofilter: series LC resonance*

Up to now we have limited our analysis to parallel combinations of nanocircuit elements for realizing simple nanofiltering systems. We recall here that, as shown analytically in [7], two nanocircuit elements may be considered in parallel when the orientation of the impinging electric field is parallel to their common interface, as it was the case in the previous scenarios. The series interconnection in this paradigm corresponds to the case in which the electric field is orthogonal to the common interface between the two nanoelements, and therefore the displacement current flows continuously from one element to the other ideally without current leakage in the surrounding circuit, as between two series elements in a standard RF circuit.

Due to the specific geometry being studied here, the parallel configuration allows better matching between the impinging wave phase fronts inside the waveguide and the nanocircuit. However, if we stack a resonant LC nanopair in series, rather than in parallel, as in Fig. 4, we may realize a stop-band nanofilter, consistent with the circuit model in the inset of Fig. 4. More generally, such series configurations may provide extra design tools for tailoring the frequency response of a complex nanocircuit. In this subsection, we analyze in detail the possibility of using a resonant series combination of nanoelements to design a stop-band nanofilter.

Figure 4 provides an example of such a resonant series configuration. A $SiO_2$ nanorod, with permittivity $\varepsilon_C = 2.16\varepsilon_0$ [22], has been stacked over a silver nanorod with the same cross-sectional dimensions: $h = h_{wg}/2 = 10\,\text{nm}$ and $d = 15\,\text{nm}$. In this way, a series resonance may "short" the waveguide and create a stop-band, expected at around 800 THz for this example. Due to the abruption of the waveguide and the strong value of the fringing fields around the interface between the two nanorods near the resonant frequency, the response of the nanofilter is slightly more complex than what the equivalent RF circuit theory would predict, but still the response in amplitude and phase approximately recalls the general trend predicted by the circuit theory (with additional resonances present due to the plasmonic resonance at the *interface* between the plasmonic and insulating materials or between the plasmonic and dielectric materials). This effect may introduce further flexibility and tunability of the frequency response of the nanofilter, due to the presence of additional localized plasmonic resonances. It is worth noting, however, that these additional stop-bands shown in Fig. 4 occur at about 50 THz below and above the main stop-band predicted by the circuit theory, and therefore allows a relatively large frequency band of operation around the main stop-band. This implies that if the desired operational band of a system is about $800 \pm 25\,THz$, for the example shown in Fig. 4, the main stop-band of the pair of nanorods will be the main one predicted by the circuit theory.

It is important to note that the field distribution at the resonance of the nanofilter, reported in Fig. 4, indeed resembles that of a resonant LC series: the electric field is oppositely oriented and very strong at the interface between the two nanocircuit elements, in order to provide a continuous displacement current flowing across the resonant pair with a zero voltage drop. This is consistent with our analytical findings in [7].

In order to better improve the agreement between the circuit theory and the full-wave simulations around the stop-band, we have simulated in Fig. 5-6 the same geometry, but with added two tiny layers ( $t = 3\,\text{nm}$ thick) of ENZ nanoinsulators on the front and the back of the nanocircuit pair and a layer of EVL connector between the resonant pair. This was done to limit the displacement current leakage at the resonance around the pair and to reduce the effect of plasmonic resonance at the interface between the two elements. The ENZ material was designed assuming a Drude model (1) with plasma frequency selected to be at the resonance frequency of the circuit, whereas the EVL material was assumed to have a permittivity $\varepsilon_{EVL} = 100\,\varepsilon_0$. (One may also use an EVL with negative permittivity, as we discuss and prove analytically for a different geometry in [9].)

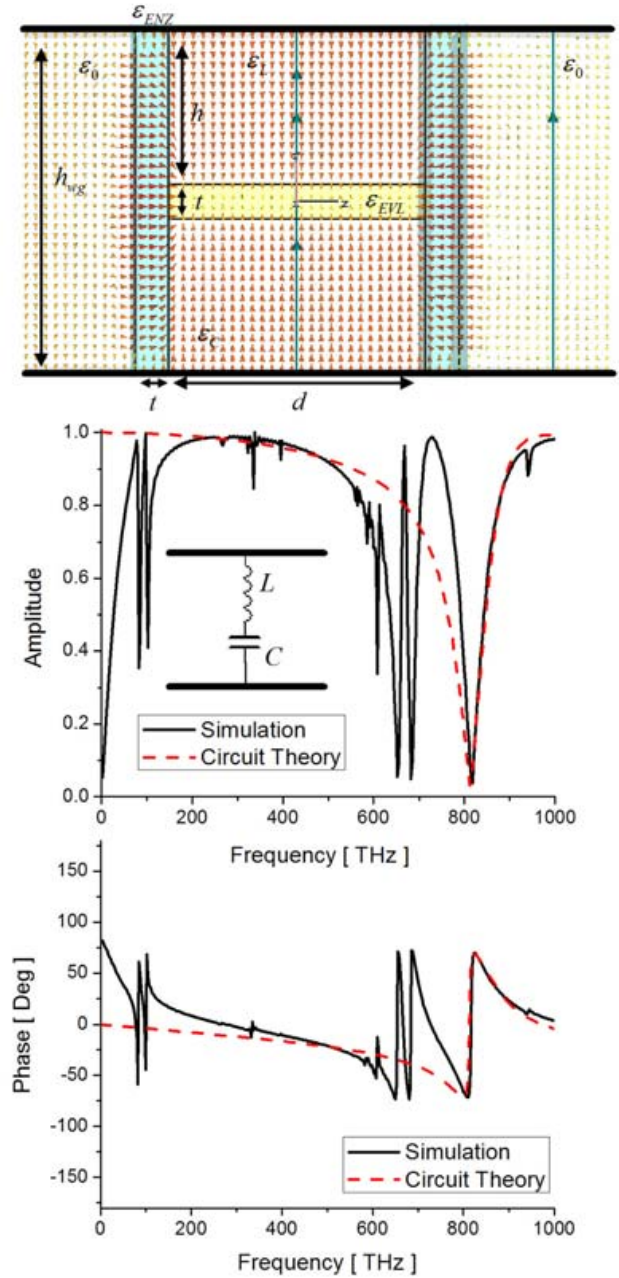

Figure 5 - Similar to Fig. 4, except here ENZ nanoinsulators and EVL nanoconnectors improve the nanocircuit functionality.

It is clear in Fig. 5 how the agreement between black (solid) and red (dashed) lines around the stop-band is better with these additions; however the intrinsic Drude dispersion of the ENZ material at lower frequencies causes the ENZ to become an epsilon-negative (ENG) layer, which consequently affects the frequency response at lower frequencies, effectively adding to the resonant pair two nanoinductors in parallel that modify the overall response of the circuit at lower frequencies. Even if in this setup the use of ENZ and EVL materials is not crucial, in more complex nanocircuits their use turns out to be important in order to avoid unwanted coupling among nanocircuit elements [6]. We note here that the ENZ and EVL material parameters employed in this simulation may arguably be hard to find in nature at the frequency of interest, but "layered" ENZ and EVL metamaterials may be straightforwardly designed by stacking together plasmonic (e.g., noble metals) and non-plasmonic (e.g., dielectric) thin-layered materials with the proper orientation with respect to the electric field, as considered in [23]. Our preliminary results and simulations in this sense confirm that it is possible to exploit this technique to fabricate artificial ENZ and EVL metamaterials with the proper frequency response for our nanocircuit applications.

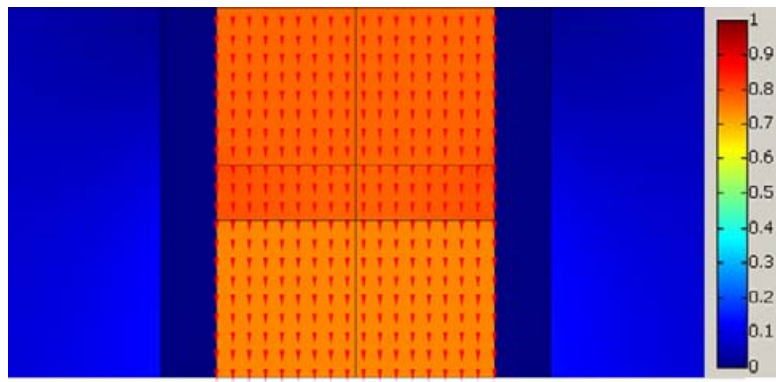

Figure 6 - Electric displacement distribution at resonance for the stop-band nanofilter of Fig. 5, simulated using finite-element software [18]. The color scale refers to the displacement vector amplitude (snapshot in time, arbitrary units), whereas the arrows refer to its direction.

The electric field and displacement distribution in Fig. 5-6, extracted at the resonance frequency of the circuit, show how the ENZ insulators allow better insulation of the series connection from the surrounding space, since the displacement current now flows all across the interface between the two nanocircuit elements, where the EVL connector is positioned. The electric field in the ENZ regions is parallel to this interface, but the low permittivity of these regions ensures that the displacement current leakage is very low. This behavior is confirmed in Fig. 6, where the displacement vector distribution $\mathbf{D} = \varepsilon\mathbf{E}$ has been evaluated for the same geometry with a finite-element software [18]. Figure 6 confirms and validates our results using a different simulation technique, and it explains more evidently the improved agreement between circuit-theory and simulations in this scenario around the resonance frequency (i.e., in the range of frequency in which the ENZ material has near-zero permittivity). It is evident how the displacement current flow is confined inside the two slabs, connected in series by an EVL material, due to the proper insulation ensured by ENZ lateral slabs. There is no electric displacement flow outside the nanofilter pair and indeed near the resonance the pair of nanoelements may be interpreted as connected in series. Once again, the additional resonances observed in this nanofilter are due to the plasmonic resonance at the interfaces between regions with oppositely-signed permittivities. Nevertheless, this nanocircuit exhibits stop-band behavior (albeit several bands), the main one of which is designed according to the analogous RF circuit theory. It is important to point out that dispersion characteristics of materials involved in such nanofilters may provide these filters with richer transfer functions that may be exploited in the overall response design.

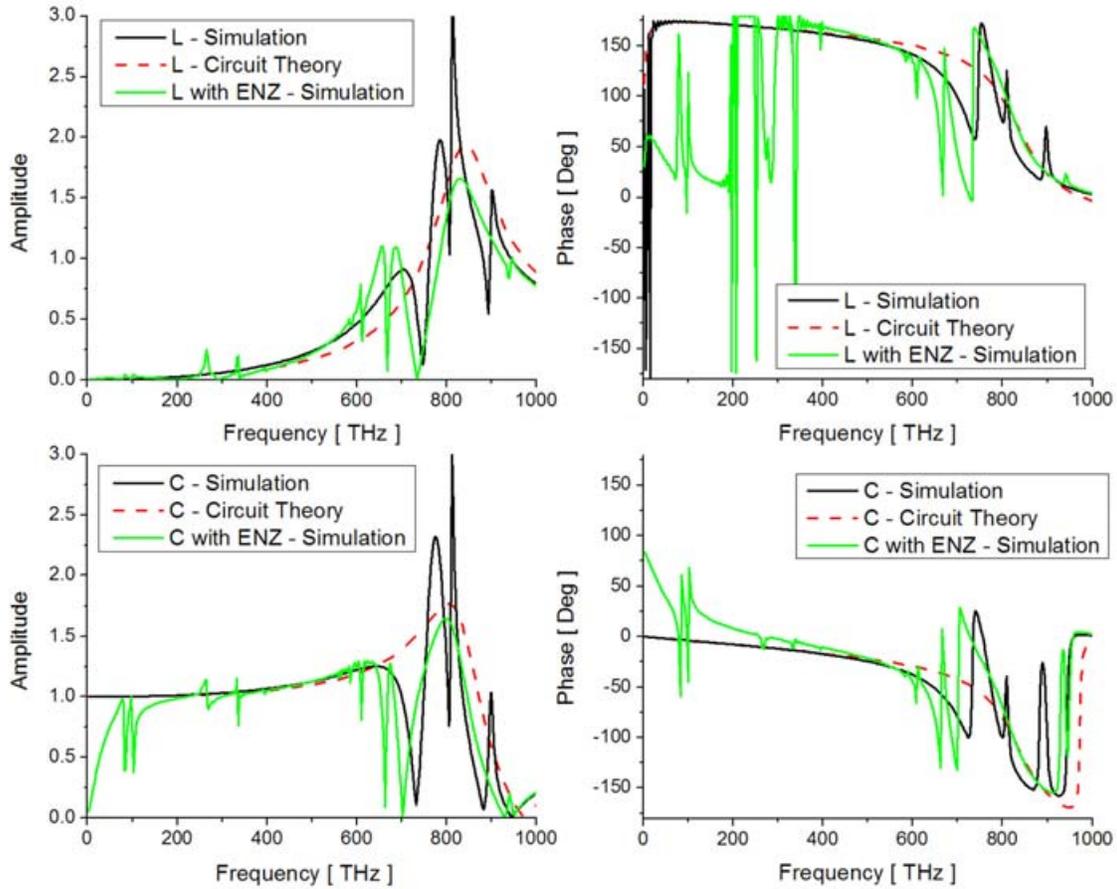

Figure 7 - Extraction of the amplitude and phase of the normalized electric field (with respect to the incident field) across the individual L and C nanoelements in the scenarios of Fig. 4 (black solid line) and Fig. 6 (green light line), compared to the ideal circ

In order to further underline the response and functionality of this stop-band series filter, Fig. 7 reports the behavior of the transfer function in this setup evaluated separately for the two nanocircuit elements, imagining to extract the signal at the two terminals of the nanoinductor (silver nanorod) and of the nanocapacitor ($SiO_2$ nanorod) separately. The plots compare the three cases of the geometry of Fig. 4 (black solid line), the geometry of Fig. 6 with nanoinsulators (green light line) and the ideal circuit response for the two elements.

There is good agreement between circuit theory and full-wave simulations, despite the abrupt change in the waveguide cross section and the dispersion of the materials. At the stop-band, even though the signal is not transmitted through the nano-filter, the two elements by themselves experience a peak in their transfer function extracted across their terminals, with 180 [deg] shift between the L and C. This is consistent with the operation of a series LC stop-band filtering device.

*e) Series combinations: Extraction of the individual element response*

Up to now we have presented several examples on how basic nanofilters might be simply realized at optical frequencies following design rules analogous to those of their low-frequency counterparts. The last example provided us with an interesting difference between a high-frequency nanocircuit and a low-frequency circuit: despite the complex resonant response of the two individual nanoelements in Fig. 4 or 6, the response measured at the output terminal of the filter shows a general stop-band response, i.e., zero transmission at the resonance frequency (as well as at other resonant frequencies). In a classic circuit, however, it may be desirable to pick the voltage/current at one of the two individual elements (i.e., L or C), similar to what reported in Fig. 7. Not having ideal "terminals" at the end of each one of the nanocircuit elements, however, it appears more difficult in this scenario to actually extract such information from the individual elements in the series combination. However, following the concepts introduced in [6]-[9] and in the previous paragraphs, we can employ a nanoconnector to extract such signal from the center of the series connection.

Fig. 8 shows the geometry that we analyze in this context: the EVL connector at the interface between the series elements of Fig. 6 has been extended down to the output terminal of the filter (here we have not employed ENZ insulators, since their presence is not crucial). In this scenario, as seen from the field distribution at the resonance, it is possible to "guide" the anti-phase field distribution present inside the nanocircuit at the resonance (as predicted by Fig. 4-7) to the end of the waveguide, analogous with the optical "shorting wire" idea we have recently presented in [9]. It is interesting to note that the electric field in the narrow EVL channel is indeed oriented longitudinally, consistent with the fact that the displacement current is being guided along the EVL "wire" to maintain the potential drop across the two elements in the empty region of the waveguide. This allows extraction of the signal from the individual elements in the series combination at the output of the filter, obtaining results that are consistent with those presented in Fig. 7. This might be a viable way to extract the transfer functions from individual optical nanocircuit elements in series.

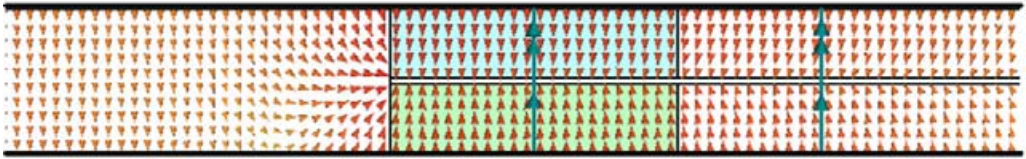

Figure 8 - Electric field distribution at the resonance for a stop-band nanofilter made by a $SiO_2$ and a silver nanorod in the series combination. An EVL nanoconnector is used to extract the optical voltage drop across each of the two nanoelements and "carry" them to the output end of the waveguide.

*f) Nanofilters at THz frequencies*

We have also been interested in applying the concepts outlined in the previous paragraphs to nanocircuits at THz frequencies. At these frequencies, some semiconductors may exhibit anomalous values for real part of permittivities. For instance, *SiC* [24], *InSb* [25] and *LiTaO$_3$* **Error! Reference source not found.** have relatively low absorption with windows of frequency over which their relative permittivities may have negative, close to zero, or very high real part. This may provide another set of materials for use in our nanofilter concepts at infrared frequencies.

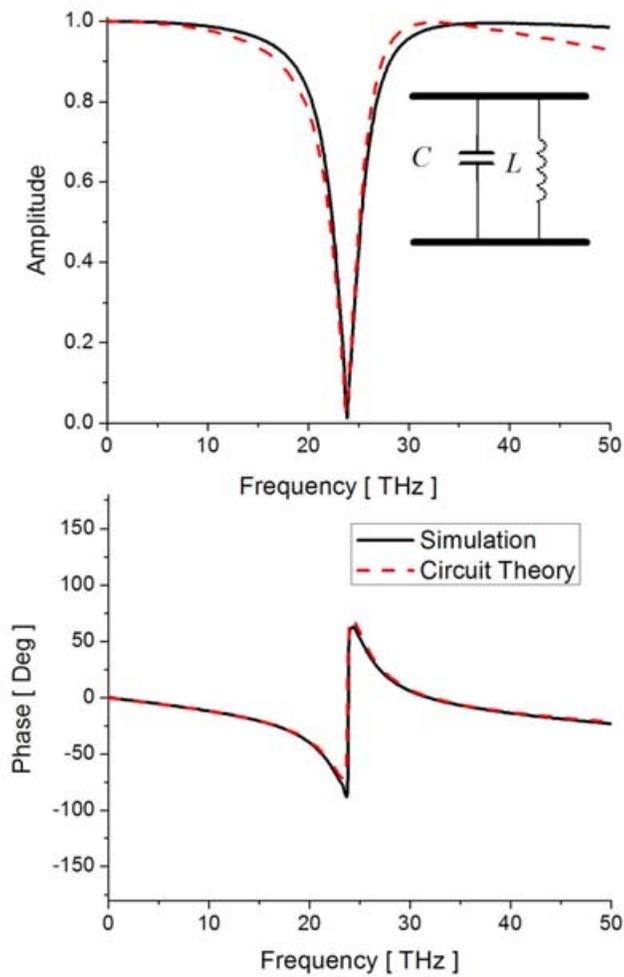

Figure 9 - Frequency response of the parallel combination of a SiC and a dielectric nanorod at THz frequencies.

Even though the previous concepts are easily scalable to lower frequencies, the main difference is in the fact that the polaritonic frequency dispersion of such materials generally follows a Lorentzian model, and therefore the bandwidth over which these concepts are applied may become narrower than those in the previous examples. As we show below, such Lorentzian dispersion may actually provide additional degrees of freedom for the nanocircuit response. Moreover, a clear advantage of this range of frequency is that it may be less challenging to realize electrically smaller nanocircuit elements at these lower frequencies.

As an example, Fig. 9 reports the transfer function for a parallel resonant pair consisting of a *SiC* nanorod of dimensions $(d = 250\,\text{nm}) \times (h = 100\,\text{nm})$ and permittivity $\varepsilon_L$ satisfying a realistic Lorentzian model with losses for *SiC* [24] in parallel with a dielectric nanorod of the same dimensions and with $\varepsilon_C = 3\varepsilon_0$, embedded in a waveguide with $h_{wg} = h$. The results are consistent with the circuit theory and show how it is possible to obtain a good stop-band nanofilter in a parallel combination in this frequency regime. This is due to the fact that the zero in transmission coefficient is obtained at the material resonance of *SiC*, around 25 THz. The parallel resonance, arising in this design around 30 THz, brings back the transfer function to unity, and for this choice of parameters has a relatively wider bandwidth of operation (consistent with the previous discussion, since now $h_{wg} = h$).

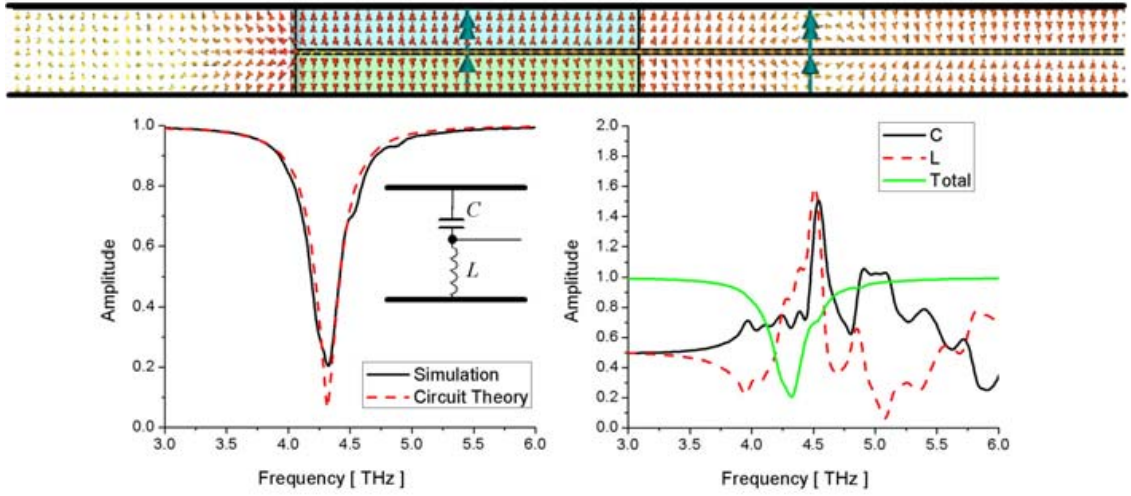

Figure 10 - Electric field distribution at the stop-band frequency, transfer function of the nanofilter, and extraction of the normalized field across the terminals of each of two nanocircuit elements for an LC series combination at THz frequencies.

Figure 10, as a further example, reports the case of a series combination obtained by combining a nanoinductor made of a Lorentzian material, with permittivity $\varepsilon_L = \left(1 + 6.7 \dfrac{\omega_L^2 - \omega_T^2}{\omega_T^2 - \omega^2 - i\omega\gamma}\right)\varepsilon_0$ with $\omega_L = 29.07\,THz$, $\omega_T = 23.79\,THz$ and $\gamma = 142.8\,GHz$ (the level of losses is analogous to those of SiC at THz frequencies), combined in series with a dielectric nanorod with $\varepsilon_C = 10\,\varepsilon_0$. The two nanocircuit elements have both height $h = h_{wg}/2 = 250\,nm$ and length $d = 2\,\mu m$ and they are separated by an EVL separator ($20\,nm$ thick) made of realistic $LiTaO_3$ material, with Lorentzian model reported in [25]. At the resonance frequency of 4.32 THz, $\varepsilon_{LiTaO_3} = (-373 + i\,433)\varepsilon_0$, acting well as an EVL material in the range of frequency of interest. The electric field distribution at the resonance shows how $LiTaO_3$, despite its non-ideal behavior with a high imaginary part, is capable of picking the resonant field distributions of two elements and guide them to the output of the waveguide, as described in the previous section. Moreover, the transfer

function shows how the stop-band behavior coincides with the expected one from the circuit theory and the extracted voltages at the end of the waveguide confirm the presence of two resonant peaks for the signals extracted from L and C separately, with a zero transmission for the sum signal, consistent with the expected nanofilter response. In this geometry, it may therefore be possible to extract separately the signals applied to L and C at the end of the waveguide. The results are in good agreement with the frequency response of a resonant LC series filter.

## 3. *Design of higher-order optical nanofilters*

After having presented the design of basic nanofilters at optical frequencies, in this section we show how even more complex configurations and frequency responses may be obtained by combining the previous concepts and having properly connected basic nanofilters forming more complex nanocircuits, similar to what is commonly done in regular low-frequency circuits.

Figure 11 refers to a third-order nano-filter, obtained by connecting in parallel a series LC filter with a nanoinductor. The two inductors are both made of silver, whereas the nanocapacitor consists of $SiO_2$. The dimensions of the parallel nanoinductor are assumed to be $(d=20\,\text{nm}) \times (h=10\,\text{nm})$, whereas the two series nanorods have size $(d/5) \times (h/2)$. Moreover, the series elements are insulated by an ENZ material with thickness $t = 4\,\text{nm}$ and a Drude model dispersion (1) with a plasma frequency selected to be around the resonant frequency of the series connection, i.e., $f_p = 810\,\text{THz}$, $\gamma = 4\,\text{THz}$.

The transfer function plots in Fig. 11 compare the circuit response for the circuit model depicted in the inset with the full-wave simulations obtained for the geometry of the figure and for the case in which the ENZ insulators are removed. The field distribution is shown at the resonant frequency of the series connection, where the zero of the filter response is expected (around $f_p$).

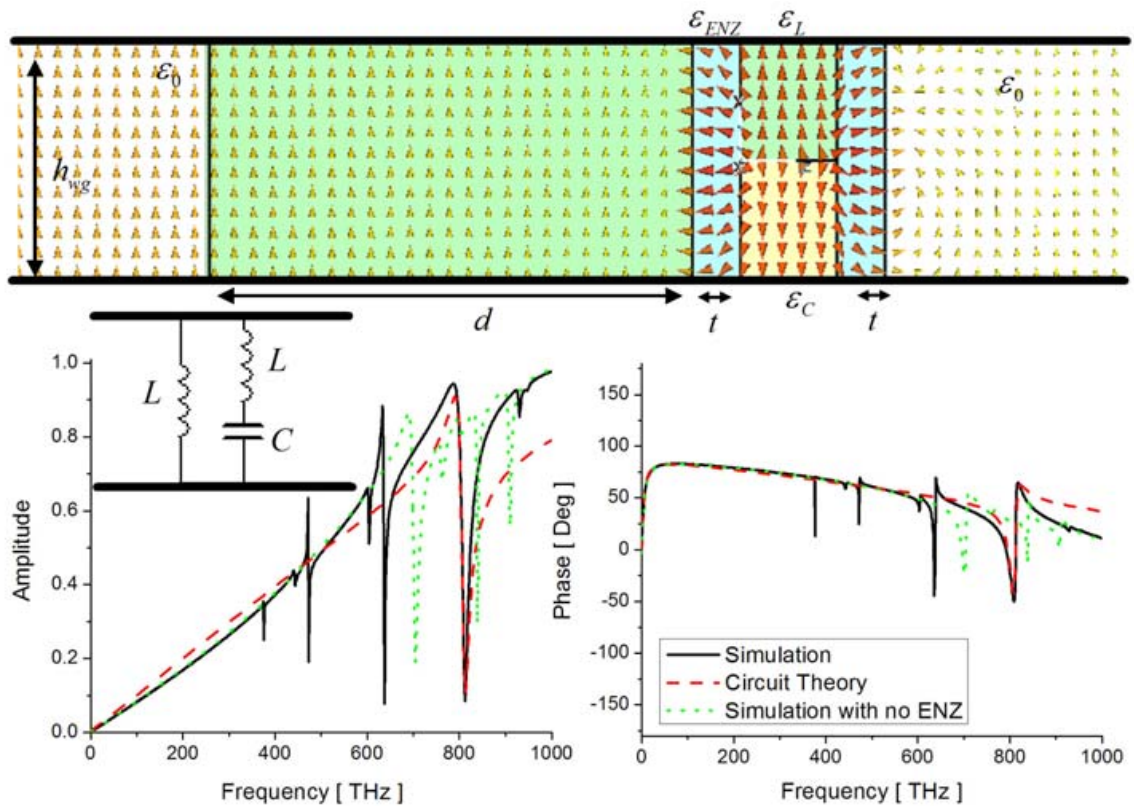

Figure 11 - Electric field distribution at the resonant frequency of the series and transfer function for a third-order nanofilter made by cascading a series LC filter with a nanoinductor. The involved materials are assumed to be silver and silicon dioxide, plus ENZ material for the nanoinsulators. The black solid line refers to the simulation when ENZ nanoinsulators are used, the dashed red line refers to ideal circuit theory simulations and the dotted green line refers to the same geometry, but without the use of ENZ nanoinsulators.

It is seen how the third-order filter combines the high-pass functionality of an RL filter with the stop-band feature of the LC series, and therefore provides further degrees of freedom for the nanocircuit response. The extracted transfer function agrees qualitatively with the ideal one obtained from circuit theory (albeit with more resonances at lower frequencies due to plasmonic resonances at interfaces), and the presence of ENZ nanoinsulators to confine the displacement current in the series connection at its resonance leads to improving the agreement around $f_p$. The electric field distribution at the series resonance, shown in Fig. 11, confirms this prediction, highlighting the out-of-phase voltage drop in the two nanocircuit elements and the absence of current leakage outside of the pair, due to the nanoinsulators.

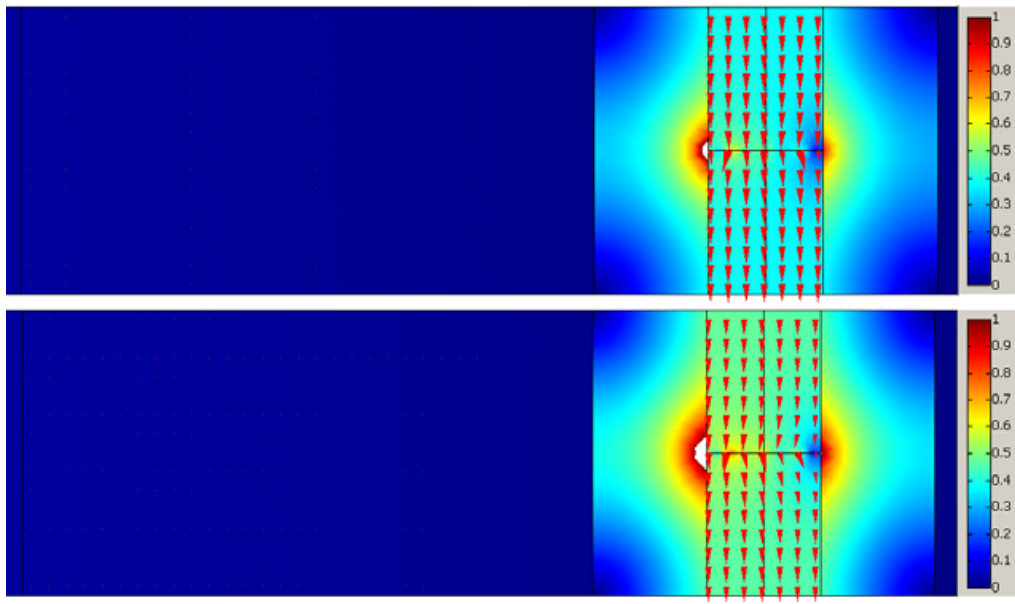

Figure 12: Electric displacement distribution for the geometry of Fig. 11 (top) and for the nanocapacitor connected in parallel with the LC structure (bottom), as evaluated using [18].

In Fig. 12, moreover, this behavior is confirmed and validated using [18], highlighting the displacement current flow at the series resonance of the nanofilter. It is noticeable how the displacement field is maximized and enhanced inside the series elements, similar to the response of a classic resonant filter, and no current leakage is allowed by the presence of the insulators.

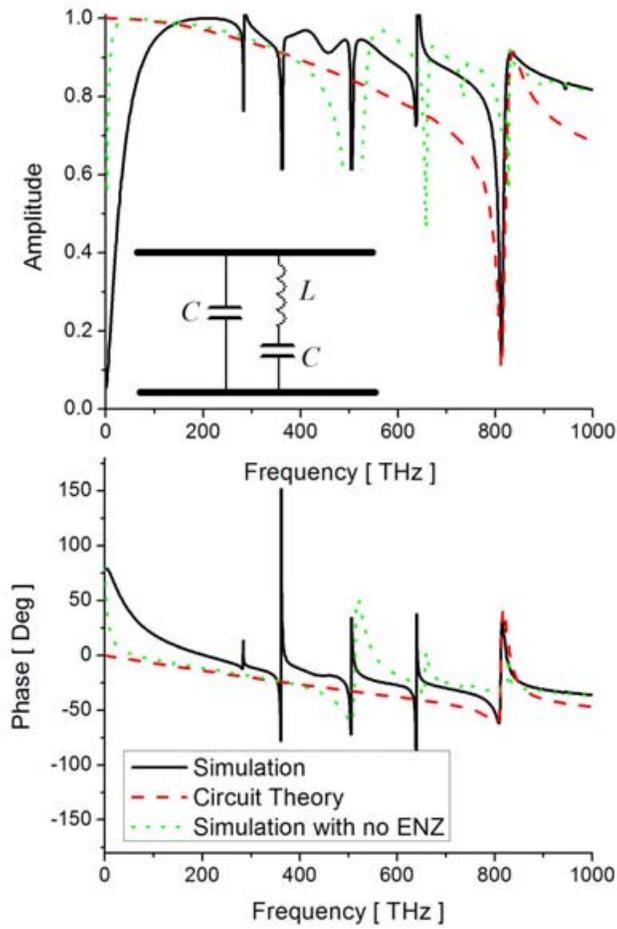

Figure 13: Transfer function for a nanofilter similar to Fig. 10, but substituting the parallel nanoinductor with a parallel nanocapacitor with $\varepsilon_C = 5\varepsilon_0$.

Figure 13 corresponds to a different third-order filter, obtained by substituting the parallel nanoinductor of Fig. 12 with a parallel nanocapacitor with $\varepsilon_C = 5\varepsilon_0$. The sizes of the elements are the same, and the functionality of the nanofilter is now expected to combine the stop-band features of the series connection with the low-pass filtering of the parallel nanocapacitor. The simulations show this to be possible and the comparison between the simulated results and the circuit response is fairly good qualitatively, considering the complexity of the nanocircuit. The presence of the insulators allows a fairly good agreement of the simulated and expected curves around the resonant frequency.

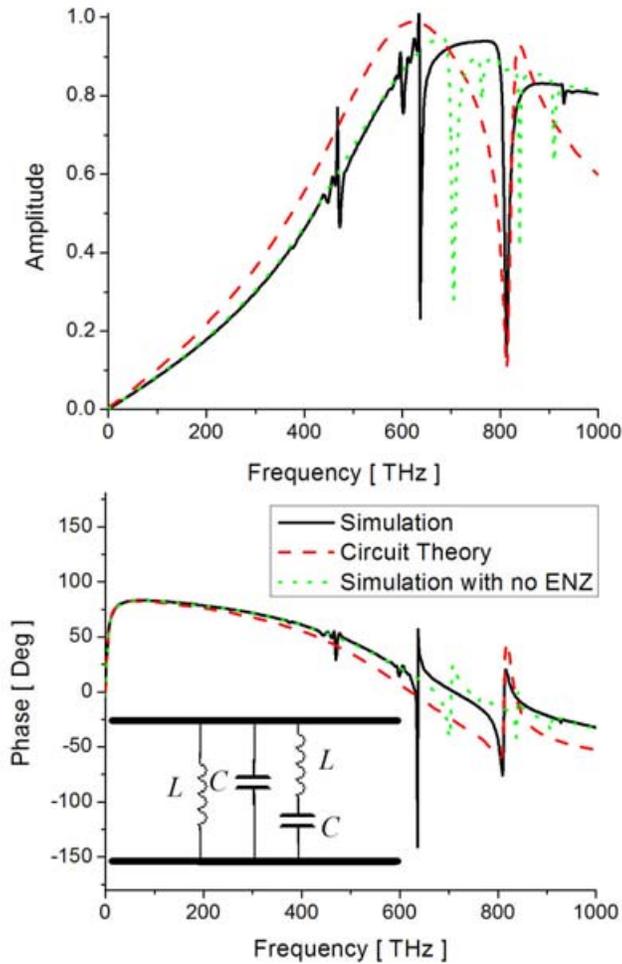

Figure 14: Electric displacement distribution (top) and transfer function (bottom) for a fourth-order nanofilter, obtained by combining in the geometry of Fig. 10-11 with both the parallel nanoinductor and nanocapacitor.

As a final example, we present in Fig. 14 the results for a fourth-order nanofilter, obtained by combining the geometries of Fig. 12 and 13 in order to have a series LC nanofilter in parallel with a parallel LC. In this case, the frequency response should give rise to two pass-bands and a stop-band, modulating a high-pass filter. This complex filter response is reasonably well obtained in the extracted response from the nanocircuit, both with and without the presence of the ENZ insulators. This final design shows how complex filtering response may be designed by utilizing plasmonic and non-plasmonic nanoparticles, suitably combined together and displaced in accordance with the impinging electric field.

## *4.*   *Conclusions*

We have reported here our theoretical investigation on the possibility of employing plasmonic and non-plasmonic nanostructures to design nanofilters and extend the functionalities of nanocircuits at infrared and optical frequencies. Relatively complex frequency responses have been obtained by employing plasmonic and dielectric nanoparticles to realize optical nanofilters with low-pass, high-pass, stop-band and pass-band functionalities. The use of nanoinsulators and nanoconnectors has been proposed to reduce unwanted coupling among the different components of the nanofilters and to extract the frequency response from series nanocircuit elements. These results may be considered an informative step towards the realization of functional nanocircuits at THz and optical frequencies.

**Acknowledgements**

This work is supported in part by the U.S. Air Force Office of Scientific Research (AFOSR) grant number FA9550-05-1-0442.

*References*

[*] To whom correspondence should be addressed, e-mail: engheta@ee.upenn.edu

[1] R. Zhia, J. A. Schuller, A. Chandran, and M. Brongersma, *Materials Today* **9**, 20 (2006).

[2] N. Engheta, A. Salandrino, and A. Alù, *Phys. Rev. Lett.* **95**, 095504 (2005).

[3] A. Alù and N. Engheta, *J. Opt. Soc. Am. B* **23**, 571 (2006).

[4] A. Alù, and N. Engheta, *Phys. Rev. B* **74**, 205436 (2006).

[5] A. Alù, and N. Engheta, *Phys. Rev. B* **75**, 024304 (2007).

[6] M. G. Silveirinha, A. Alù, J. Li, and N. Engheta, "Nanoinsulators and nanoconnectors for optical nanocircuits," under review, online at: http://arxiv.org/abs/cond-mat/0703600.

[7] A. Alù, A. Salandrino, and N. Engheta, "Parallel, Series, and Intermediate Interconnections of Optical Nanocircuit Elements - Part 2: Nanocircuit and Physical Interpretation," submitted to J. Opt. Soc Am. B, online at: http://arxiv.org/abs/0707.1003.

[8] A. Alù, A. Salandrino, and N. Engheta, "Coupling of Optical Lumped Nanocircuit Elements and Effects of Substrates," submitted to Opt. Expr., online at: http://arxiv.org/abs/0706.1316.

[9] A. Alù, and N. Engheta, "Optical 'Shorting Wires'," Opt. Expr., in press, online at: http://arxiv.org/abs/0706.4120.

[10] R. W. Rendell, and D. J. Scalapino, *Phys. Rev. B* **24**, 3276 (1981).

[11] X. C. Zeng, P. M. Hui, D. J. Bergman, and D. Stroud, *Phys. Rev. B* **39**, 13224 (1989).

[12] D. A. Genov, A. K. Sarychev, V. M. Shalaev, and A. Wei, *Nano Lett.* **4**, 153 (2004).

[13] A. I. Csurgay, and W. Porod, *Int. J. Circuit Theory and Applications* **32**, 339 (2004).

[14] L. D. Landau, and E. M. Lifshitz, *Electrodynamics of Continuous Media, Course of Theoretical Physics*, vol.8, (Oxford, Elsevier Butterworth-Heinemann, 2004).

[15] C. F. Bohren, and D. R. Huffman, *Absorption and Scattering of Light by Small Particles* (Wiley, New York, 1983).

[16] A. S. Sedra, and K. C. Smith, *Microelectronic Circuits* (Oxford University Press, USA, 2003).

[17] CST Design Studio$^{TM}$ 2006B, www.cst.com.

[18] COMSOL Multiphysics 3.2 (2006), COMSOL Inc., www.comsol.com

[19] G. D. Mahan, and G. Obermair, *Phys. Rev.* **183**, 834 (1969).

[20] S. Adachi, *Phys. Rev. B* **38**, 12966 (1988).

[21] P. B. Johnson, and R. W. Christy, *Phys. Rev. B* **6**, 4370 (1972).

[22] I. H. Malitson, *J. Opt. Soc. Am.* **55**, 1205 (1965).

[23] S. A. Ramakrishna, J. B. Pendry, M. C. K. Wiltshire, and W. J. Stewart, *J. Mod. Opt.* **50**, 1419 (2003).

[24] W.G. Spitzer, D. Kleinman, and J. Walsh, *Phys. Rev.* **113**, 127, (1959).

[25] J. Gómez Rivas, C. Janke, P. Bolivar, and H. Kurz, *Opt. Express* **13**, 847 (2005).

[26]  M. S. Wheeler, J. S. Aitchison, and M. Mojahedi, *Phys. Rev. B* **72**, 193103 (2005).